\documentclass[preprint,12pt]{elsarticle}

\usepackage{amsmath}

\usepackage{amssymb}

\newcommand{\be}{\begin{equation}}
\newcommand{\ee}{\end{equation}}
\newcommand{\bea}{\begin{eqnarray}}
\newcommand{\eea}{\end{eqnarray}}

\begin{document}

\begin{frontmatter}



\title{Emergence of simple characteristics for heterogeneous complex social agents}


\author{Eric Bertin}

\address{Univ.~Grenoble Alpes, CNRS, LIPHY, F-38000 Grenoble, France}

\begin{abstract}
Models of interacting social agents often represent agents as very simple entities having a small number of degrees of freedom, as exemplified by binary opinion models for instance. Understanding how such simple individual characteristics may emerge from potentially much more complex agents is thus a natural question.
It has been proposed recently in [E. Bertin, P. Jensen, C.~R. Phys. {\bf 20}, 329 (2019)] that some types of interactions among agents with many internal degrees of freedom may lead to a `simplification' of agents, which are then effectively described by a small number of internal degrees of freedom.
Here, we generalize the model to account for agents intrinsic heterogeneity.
We find two different simplification regimes, one dominated by interactions, where agents become simple and identical as in the homogeneous model, and one where agents remain strongly heterogeneous although effectively having simple characteristics.
\end{abstract}

\begin{keyword}
Sociophysics, statistical physics, heterogeneity, phase transition.
\end{keyword}

\end{frontmatter}

\section{Introduction}

When modeling complex systems, statistical physicists often posit that the interacting entities they consider have simple individual properties, and that the possibly more complex behavior observed at a collective level, when considering many such simple entities, is simply the result of interactions between entities \cite{Castellano,LesHouches,BarratBook,BouchaudJSP13}. The emergence of macroscopically different properties from simpler interacting entities, often through a collective symmetry breaking mechanism, has been emphasized long ago by P.~W. Anderson in his seminal paper "More is Different" \cite{Anderson} as one of the key mechanisms at play to account for the wealth of different objects or behaviors found in the real world.
In this paper, Anderson actually presents the statistical physics approach as a generic and somewhat iterative method, to deal with the emerging complexity in a multilevel way, the outcome of a given level of analysis being the building blocks (i.e., the interacting entities) of the next level.
The key role played by symmetry breaking phenomena in the emerging complexity at collective scale has been widely acknowledged in many different contexts (see, e.g., \cite{Chaikin,LeBellac,LesHouches,activematterRMP} among many others).
Interestingly, a perhaps less explicit suggestion of Anderson's paper is also to consider statistical physics models where the interacting entities are not simple objects with a handful of characteristics, but are already rather complex objects with many internal degrees of freedom.
Although it has been formulated almost fifty years ago, this suggestion looks very timely by now.

In the last decades, statistical physics has gone beyond the equilibrium paradigm based on molecular entities, and has indeed started to consider assemblies of macroscopic and potentially complex objects like grains of sand \cite{deGennesRMP,PuglisiBook}, or active particles modeling for instance some types of bacteria, or self-phoretic colloids \cite{activematterRMP}.
However, in these examples, macroscopic particles may in practice still be considered as simple particles, as their many internal degrees of freedom can be subsumed into a small number of effective parameters encoding their macroscopic, non-equilibrium character, like dissipation coefficients \cite{deGennesRMP,PuglisiBook} or self-propulsion forces \cite{activematterRMP}. To find genuine examples of assemblies of complex entities in a theoretical context, one may rather turn to the field of population dynamics and evolution in theoretical biology, where for instance large populations of individuals characterized by a complicated genome evolve under some evolutionary rules \cite{DrosselAdvPhys,SellaPNAS}.

When considering models of social systems, it would be necessary at first sight to take into account the intrinsic complexity of human beings \cite{Castellano}. However, this complexity is too extreme to be captured by any type of statistical physics models, so that statistical physicists have often considered very simple models of social agents, retaining a small number of characteristics, for instance a binary \cite{Sznajd} or continuous \cite{Deffuant} opinion.
A natural question is then to understand how more complex agents could in some situations reduce their intrinsic complexity to effectively appear as simple. This question can already be addressed within the framework of statistical physics, because there is no need to model the full complexity of human beings to address at least some aspects of this issue.
A simple tentative answer has been given in \cite{Jensen2019}, by illustrating on a toy model how the simplification of agents with many internal degrees of freedom may result from interactions among agents. This point of view is qualitatively consistent with the idea advocated by some sociologists that `the whole is less than the parts' \cite{Latour,Jensen-politics}, in the sense that, roughly speaking, human beings may leave aside part of their complexity to build a group.
It has been argued in \cite{Jensen2019} that this point of view can be reconciled with the seemingly antagonist viewpoint of statistical physicists according to which `the whole is more than the sum of the part', due to collective phenomena and symmetry breakings.

Yet, the model introduced in \cite{Jensen2019} considered an assembly of identical agents, while agents heterogeneity may be expected to be an important characteristics of human beings.
In this note, we extend the model of \cite{Jensen2019} by including heterogeneity between agents.
We show, using methods inspired by the physics of glasses, that two different types of agents simplification can occur in this model, one driven by interactions as in \cite{Jensen2019}, and the other one driven by heterogeneity.

\section{Model}

We introduce a model of complex agents generalizing the one introduced in \cite{Jensen2019} by now considering heterogeneous agents.
The model is composed of $N$ interacting agents having an internal state described by a configuration
$\mathcal{C}_i \in \{1,...,H\}$, with $i=1,\dots,N$ and where the number $H$ of configurations is large. 
We write for later convenience $H$ in the form $H=n_0^M$, where $n_0$ is a fixed integer number, and $M$ is assumed to be large. This is typically the case if the configuration $\mathcal{C}_i$ is composed of $M$ degrees of freedom, each of which taking $n_0$ possible values.
Each agent is endowed with a characteristic that can be either present or absent depending on the configuration 
$\mathcal{C}_i$. Intuitively, this characteristic could be a preference for a specific kind of music or tempo for the members of a vocal ensemble \cite{Jensen-politics} for instance, or more generally be related to a binary opinion \cite{Sznajd}. This feature is conveniently encoded by a variable $S_i(\mathcal{C}_i) \in \{0,1\}$: the characteristic is present when $S_i(\mathcal{C}_i)=1$, and absent when $S_i(\mathcal{C}_i)=0$.
The general idea is that the characteristic would be present only in a small number of internal states, so that it would typically remain unobserved except if there is a strong probability bias towards the few configurations for which $S_i(\mathcal{C}_i)=1$.
To implement this idea in practice in the model, we assume that the characteristic is present in a single configuration, that we label $\mathcal{C}_i=1$.

We now need to define the dynamics of agents. Following standard practice in the modeling of social agents \cite{Nadal04}, we assume that the dynamics is driven by an individual utility function
$u_i = u_i(\mathcal{C}_i\vert \mathcal{C}_{j\ne i})$ that accounts both for individual preferences and for interactions with agents.
An agent $i$ stochastically changes configurations according to the following rule.
Given the current configuration $\mathcal{C}_i$, the new configuration $\mathcal{C}'_i$ is randomly chosen with a probability rate given by the logit rule,
\be \label{eq:transition:rate}
W(\mathcal{C}'|\mathcal{C}) = \frac{1}{1+e^{-\Delta u_i/T}}
\ee
with $\Delta u_i = u_i(\mathcal{C}'_i) - u_i(\mathcal{C}_i)$ the variation of utility generated by the change of configuration (note that configurations $\mathcal{C}_j$ of the other agents $j \ne i$ are kept fixed).
The parameter $T$ plays a role similar to temperature in statistical physics, and characterizes the degree of stochasticity in the decision rule.

Our goal is to model heterogeneous agents that interact through their characteristic $S_i$ (their internal state is otherwise invisible to other agents).
With this aim in mind, we choose the following form of the utility function of agent $i$,
\be \label{eq:def:utility}
u_i(\mathcal{C}_i\vert \mathcal{C}_{j\ne i}) = U_i(\mathcal{C}_i)
+ \frac{K}{N} \sum_{j (\ne i)} S_i(\mathcal{C}_i) S_j(\mathcal{C}_j) \,,
\ee
where $U_i(\mathcal{C}_i)$ is the intrinsic (or idiosyncratic) utility of configuration $\mathcal{C}_i$ for agent $i$, and $K$ is the coupling constant characterizing the interaction with the other agents. Note the $1/N$ scaling of the interaction term, typical of fully connected models where all particles or agents interact with each other in a similar way.
This interaction terms was already present in the homogeneous model of Ref.~\cite{Jensen2019}.
We then model the heterogeneity of agents as a quenched randomness of the intrinsic utilities (which were absent from the model of \cite{Jensen2019}).
More precisely, for all $i=1,\dots,N$ and $\mathcal{C}_i=1,\dots,H$, the intrinsic utility $U_i(\mathcal{C}_i)$ is randomly drawn from a Gaussian distribution $\rho(U)$,
\be \label{eq:def:rhoU}
\rho(U) = \frac{1}{\sqrt{\pi M J^2}} \, e^{-U^2/MJ^2} \,,
\ee
where we recall that $M$ is defined by $H=n_0^M$.
The utilities $U_i(\mathcal{C}_i)$ do not change in time.

The utility variation $\Delta u_i $ can be reformutated as the variation $\Delta E = \Delta u_i $ of a global observable $E$ that plays a role similar to the energy in physics (up to a change of sign). Note that contrary to $u_i$, the pseudo-energy $E$ does not depend on the agent $i$. 

Here, the function $E$ takes the form
\be \label{eq:energy:model1}
E(\mathcal{C}_1,\dots,\mathcal{C}_N) = \sum_{i=1}^N U_i(\mathcal{C}_i) + \frac{K}{2N} \sum_{i,j (i\ne j)} S_i(\mathcal{C}_i) S_j (\mathcal{C}_j)\,.
\ee
The quantity $E$ is thus different from the total utility $\sum_i u_i$, due to the factor $\frac{1}{2}$ in the interaction term.
Note also that the present model shares similarities with the Random Energy Model \cite{DerridaPRL,DerridaPRB,BouchMez97}, as well as with the Ising model \cite{LeBellac}, the Potts model \cite{review-Potts} or the Blume-Emery-Griffiths spin-1 model \cite{BEG71}. However, it also exhibits important differences with each of these models.

Given the property $\Delta u_i = \Delta E$, the dynamics defined by the transition rate Eq.~(\ref{eq:transition:rate}) obeys the detailed balance property in terms of the equilibrium distribution
\be \label{eq:equil:dist}
P(\mathcal{C}_1,\dots,\mathcal{C}_N) = \frac{1}{Z}\, e^{\beta E(\mathcal{C}_1,\dots,\mathcal{C}_N)} \,,
\ee
with $\beta \equiv 1/T$, and where $Z$ is a normalization constant.
Now that we determined the equilibrium distribution of the model, our goal is to investigate its phase diagram to assess the effect of the competition between agents heterogeneity (due to their quenched intrinsic utility) and collective effects that could arise from interactions.
As recalled in the introduction, the homogeneous version of the model, studied in \cite{Jensen2019}, exhibits a transition driven by interactions between a phase where agents essentially visit all their internal states and thus have no strongly preferred configurations, and an ordered state where all agents `standardize' in the same configuration $\mathcal{C}_1$ such that $S_i=1$ (i.e., the characteristic of the agent becomes visible).
In the present generalization of the model, we wish to explore whether the standardized state survives the heterogeneity of agents intrinsic utility. Indeed, the internal state $\mathcal{C}_1$ such that $S_i=1$ may have a  lower intrinsic utility $U_i(\mathcal{C}_1)$ than other more favored configurations of agent $i$, and this may prevent a common standardization of all agents in the same state.

To investigate this issue, we introduce the order parameter $q$ defined as
\be
q = \frac{1}{N} \sum_{i=1}^N S_i\,.
\ee
If the number $N$ of agents is large, the pseudo-energy $E$ can be expressed in terms of the order parameter $q$ as
\be \label{eq:expr:E}
E = \sum_{i=1}^N U_i +  \frac{1}{2} N K q^2 \,.
\ee
We now wish to determine the distribution $P(q)$ of the order parameter $q$. This distribution is obtained by summing the joint distribution $P(\mathcal{C}_1,\dots,\mathcal{C}_N)$ over all configurations $(\mathcal{C}_1,\dots,\mathcal{C}_N)$ having a given value of $q$, namely
\be \label{eq:def:Pq}
P(q) = \sum_{\mathcal{C}_1,\dots,\mathcal{C}_N} P(\mathcal{C}_1,\dots,\mathcal{C}_N)
\, \delta \Big( q(\mathcal{C}_1,\dots,\mathcal{C}_N) - q\Big)
\ee
with $\delta$ the Kronecker delta.
For a given value of $q$, there are $qN$ agents in configuration $\mathcal{C}_i=1$ and $(1-q)N$ agents in any of the other $n_0^M-1$ configurations.
We denote by $\mathcal{S}_{q,N}$ a subset with $qN$ elements of the set $[1,N]$.
With these notations, $P(q)$ can be written as
\be \label{eq:Pq:v2}
P(q) = \frac{1}{Z} \, \sum_{\mathcal{S}_{q,N}}
\left[ \prod_{i\in \mathcal{S}_{q,N}} e^{\beta U_i(1)} \right]
\left[ \prod_{i\notin \mathcal{S}_{q,N}} \sum_{\mathcal{C}_i=2}^{n_0^M} e^{\beta U_i(\mathcal{C}_i)} \right]
\, e^{\frac{1}{2}\beta N K q^2} \,,
\ee
where in the second product, the index $i$ is implicitly restricted to the interval $[1,N]$.
The distribution $P(q)$ defined in Eq.~(\ref{eq:def:Pq}) actually depends on the specific realization of the random utilities $U_i(\mathcal{C}_i)$, so that $P(q)$ should in principle be eventually averaged over these random utilities.
However, to make calculations easier, we do not compute explicitly the average over $U_i(\mathcal{C}_i)$, but rather use heuristic arguments to evaluate the typical values of the random quantities appearing in Eq.~(\ref{eq:Pq:v2}).
Such an estimate is expected to be sufficient to determine the leading exponential behavior $P(q) \sim e^{-Nf(q)}$ of the distribution $P(q)$.
The first product between brackets in Eq.~(\ref{eq:Pq:v2}) can be rewritten as the exponential of $\sum_{i\in \mathcal{S}_{q,N}} \beta U_i(1)$. The latter sum is (up to the factor $\beta$) a sum of independent and identically distributed Gaussian random variables drawn from the distribution Eq.~(\ref{eq:def:rhoU}). The sum is thus also Gaussian distributed, with zero mean and variance $\frac{1}{2} qNm(\beta J)^2$. It follows that
\be
\prod_{i\in \mathcal{S}_{q,N}} e^{\beta U_i(1)} \sim e^{O(\sqrt{N})}\,,
\ee
and this product can thus be neglected (i.e., replaced by $1$) when looking for the behavior of $P(q)$ at exponential order in $N$.

\begin{figure}[t]
\centering\includegraphics[width=0.7\linewidth]{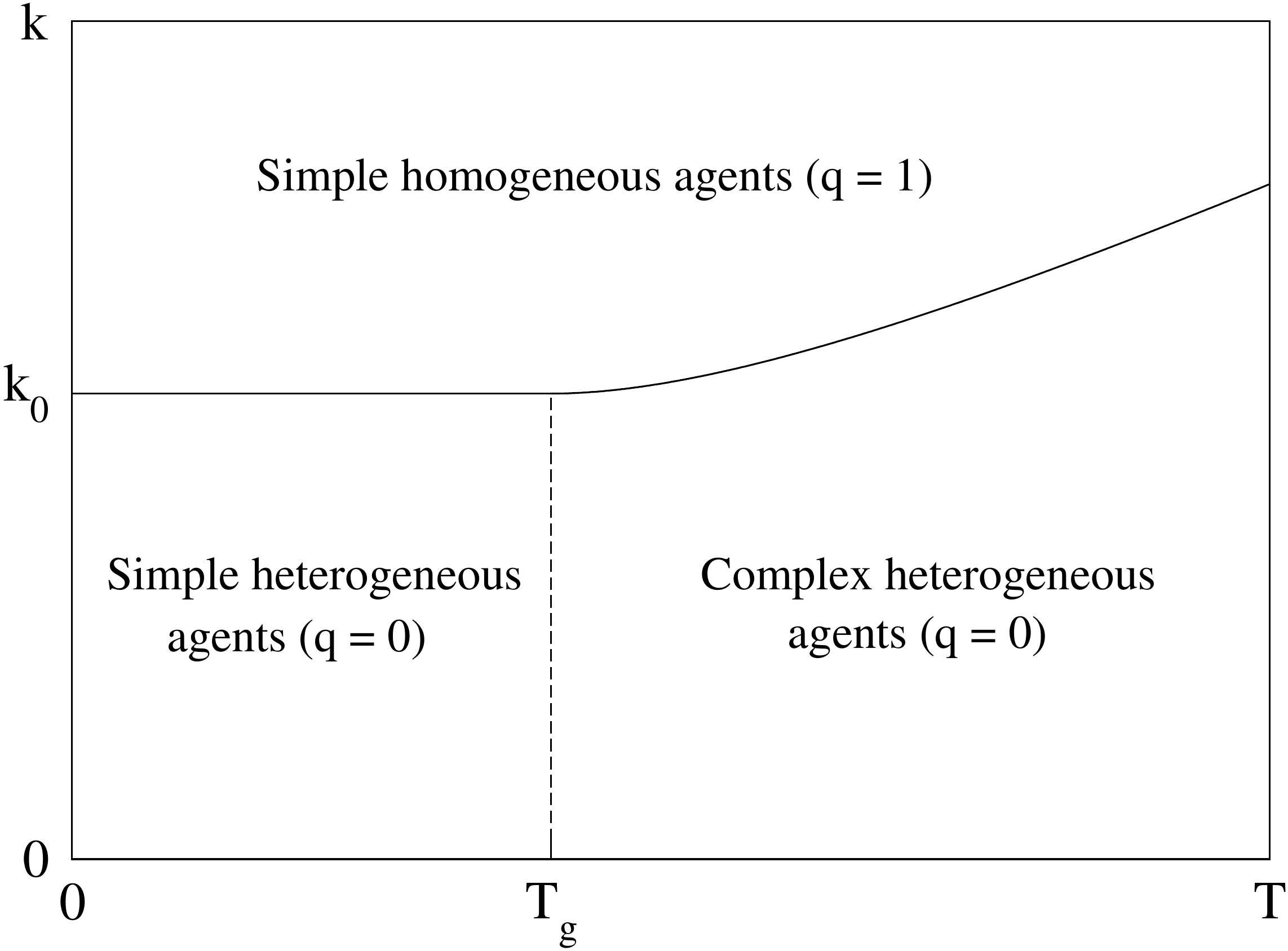}
\caption{Phase diagram of the model in the $(k,T)$-plane (reduced coupling constant $k=K/M$ versus temperature), showing the three different regions separated by the critical line $k_c(T)$ (full line) and the `glass' temperature $T_{\rm g}$ (dashed vertical line).
For $k<k_c(T)$, agents configurations do not overlap ($q=0$), while for $k>k_c(T)$ agents become standardized by interactions ($q=1$).
The $q=0$ area is subdivided into high- and low-temperature regions. For $T>T_{\rm g}$, agents may be in any of their internal configurations, while for $T<T_{\rm g}$, they are dynamically blocked in the few configurations with the highest intrinsic utility. In this latter case, agents appear simple but remain heterogeneous.}
\label{fig:phasediag}
\end{figure}

The key point in order to determine $P(q)$ explicitly is now to evaluate the typical value $\mathcal{Z}_{\rm typ}$ of the sum 
\be
\mathcal{Z}_i = \sum_{\mathcal{C}_i=2}^{n_0^M} e^{\beta U_i(\mathcal{C}_i)}\,,
\ee
where we recall that each $U_i(\mathcal{C}_i)$ is an independent random variable drawn from the distribution $\rho(U)$ given in Eq.~(\ref{eq:def:rhoU}).
Once this estimate is known, the distribution $P(q)$ can be approximated as
\be \label{eq:Pq:v3}
P(q) \sim \frac{1}{Z} \,\binom{N}{qN}\, \mathcal{Z}_{\rm typ}^{(1-q)N} \, e^{\frac{1}{2}\beta N K q^2} \,.
\ee
The quantity $\mathcal{Z}_i$ has the same form as the partition function of the Random Energy Model (REM) \cite{DerridaPRL,DerridaPRB}, and we can thus borrow some methods from the REM to evaluate it.
A standard approach to study the REM is to evaluate the density of states of a typical realization of the disorder.
It is convenient at this stage to define the rescaled variable $u=U/M$.
Denoting as $n(u)$ the density of state of a given realization, the average $\langle n(u) \rangle$ over the disorder is given by
\be
\langle n(u) \rangle = n_0^M \, \tilde{\rho}(u) \sim e^{M(\ln n_0 - u^2/J^2)}
\ee
to exponential order in $M$, having defined $\tilde{\rho}(u) = M \rho(Mu)$.
Hence if $\ln n_0 - u^2/J^2 >0$, corresponding to $|u| < u_0 = J \sqrt{\ln n_0}$, the average density of state $\langle n(u) \rangle$ is exponentially large with $M$, and for a typical sample $n(u) \approx \langle n(u) \rangle$.
By contrast, when $|u| > u_0$, the average density of state is exponentially small with $M$, meaning that in most realizations there are actually no states with $|u| > u_0$ (the exponentially small value of the average density of states comes from very rare and atypical realizations having a few states in this range).
Therefore, to describe a typical realization, one can in practice consider that the density of state is equal to zero for 
$|u| > u_0$, and equal to $\langle n(u) \rangle$ for $|u| < u_0$.
We can thus evaluate $\mathcal{Z}_i$ as
\be \label{eq:Ztyp}
\mathcal{Z}_i = \sum_{\mathcal{C}_i} e^{\beta U_i(\mathcal{C}_i)} \approx \sqrt{\frac{M}{\pi J}}
\int_{-u_0}^{u_0} du \, e^{M(\ln n_0 - u^2/J^2 + \beta u)} \equiv \mathcal{Z}_{\rm typ}\,.
\ee
The integral in Eq.~(\ref{eq:Ztyp}) can be evaluated by a saddle-point calculation in the large $M$ limit.
Defining $g(u)=\ln n_0 - u^2/J^2 + \beta u$, $\mathcal{Z}_{\rm typ}$ is given for large $M$ by
$\mathcal{Z}_{\rm typ} \sim e^{Mg_{\max}}$, where $g_{\max}$ is the maximum value of $g(u)$ over the interval
$[-u_0,u_0]$. Let us first look for the maximum of $g(u)$ over the whole real axis.
Defining $u^*$ such that $g'(u^*)=0$, we find $u^*=\frac{1}{2}\beta J^2$.
Recalling that $\beta=1/T$, and defining
\be
T_{\rm g} = \frac{J}{2\sqrt{\ln n_0}} \,,
\ee
we find that $-u_0<u^*<u_0$ for $T>T_{\rm g}$, whereas $u^* \ge u_0$ for $T \le T_{\rm g}$.
Hence $g_{\max} = g(u^*)$ for $T>T_{\rm g}$, and $g_{\max} = g(u_0)$ for $T\le T_{\rm g}$, taking into account the fact that $g(u)$ is monotonously increasing for $u<u^*$.
We thus obtain
\be \label{eq:Ztyp:explicit}
\mathcal{Z}_{\rm typ} \approx 
\begin{cases}
e^{M(\ln n_0 + \frac{1}{4}\beta^2 J^2)} & \qquad \mbox{if} \  T>T_{\rm g} \,,\\
e^{M\beta J \sqrt{\ln n_0}}  & \qquad  \mbox{if} \  T \le T_{\rm g} \,.
\end{cases}
\ee
Using Eqs.~(\ref{eq:Pq:v3}) and (\ref{eq:Ztyp:explicit}), we obtain
after expanding the factorials thanks to the Stirling formula that
$P(q)$ takes a large deviation form $P(q) \sim e^{Nf(q)}$.
In physical terms, $f(q)$ may be thought of as (the opposite of) a free energy. 
The explicit expression of $f(q)$ depends on the temperature range.
For $T>T_{\rm g}$, $f(q)$ reads
\be
f(q) =- q \ln q - (1-q) \ln (1-q)
+ (1-q) M \left[ \ln n_0 + \frac{J^2}{4T^2} \right] + \frac{1}{2T} K q^2 \,,
\label{eq:fq:TgtTg}
\ee
whereas for $T\le T_{\rm g}$,
\be
f(q) = - q \ln q - (1-q) \ln (1-q)
+ (1-q)M \, \frac{J}{T} \sqrt{\ln n_0} + \frac{1}{2T} K q^2 \,.
\label{eq:fq:TltTg}
\ee
Note that we did not take into account in $f(q)$ the contribution coming from the normalization factor $Z$, as this would simply add a constant to $f(q)$.
Inspired by Ref.~\cite{Jensen2019}, where interesting results were obtained for a coupling constant $K \sim M$,
we assume in what follows that 
\be
K = k M\,,
\ee
and take the reduced constant $k$ as the relevant control parameter in the model (on top of temperature $T$).
For large $M$, the expression of $f(q)$ then simplifies to
\be
f(q) = 
\begin{cases}
\frac{M}{T} \left[ (1-q) \left( T\ln n_0 + \frac{J^2}{4T} \right) + \frac{1}{2} k q^2 \right]  & \quad \mbox{if} \ T>T_{\rm g} \,,\\
\frac{M}{T} \left[  (1-q) \, J \sqrt{\ln n_0} + \frac{1}{2} k q^2 \right] & \quad \mbox{if} \ T\le T_{\rm g} \,.
\end{cases}
\ee
We first observe that in this large-$M$ approximation, $f(q)$ is a convex function of $q$ for all values of temperature $T$, so that the maximum of $f(q)$ over the interval $0\le q\le 1$ is either $f(0)$ or $f(1)$.
The most probable state is found to be $q=0$ for $k<k_{\rm c}(T)$ and $q=1$ for $k>k_{\rm c}(T)$, where
the critical line $k_{\rm c}(T)$ is defined as (with $k_0=2J \sqrt{\ln n_0}$)
\be
k_{\rm c}(T) =
\begin{cases}
\frac{k_0}{2} \left(\frac{T}{T_{\rm g}}+\frac{T_{\rm g}}{T}\right) & \quad \mbox{if} \ T>T_{\rm g} \,, \\
k_0  & \quad \mbox{if} \ T\le  T_{\rm g} \,.
\end{cases}
\ee
The curve $k_{\rm c}(T)$ thus separates the $(k,T)$ phase diagram into two regions, a region with $q=0$ at low coupling and a region with $q=1$ at high coupling.
The corresponding phase diagram is plotted in Fig.~\ref{fig:phasediag}.
Note that for $J=0$ (i.e., in the absence of disorder), the glassy region in the phase diagram disappears, and one recovers the phase transition at $T_{\rm c} = k/(2\ln n_0)$ between a high-temperature phase with $q=0$ and a low-temperature phase with $q=1$ found in the homogeneous model \cite{Jensen2019}.

Besides, we have also seen that a change of behavior occurs at $T=T_{\rm g}$.
For $T>T_{\rm g}$, the agents dynamically visit a large number of configurations, while for $T<T_{\rm g}$ their dynamics becomes essentially frozen, and only few configurations have a significant probability to be visited.
In other words, for $T<T_{\rm g}$ agents become `stuck' in a small number of configurations having the highest utility.
In the context of the Random Energy Model for glasses, the temperature $T_{\rm g}$ corresponds to the glass transition.

Hence there are actually three different regions in the phase diagram
shown in Fig.~\ref{fig:phasediag}.
For $T>T_{\rm g}$ and $k<k_{\rm c}(T)$, agents have no preferred configurations and visit many different configurations over time.
For $T<T_{\rm g}$ and $k<k_{\rm c}(T_{\rm g})$, each agent spends a lot of time in a small set of preferred configurations. In other words agents look simple, but they remain different one from the other.
This regime is dominated by agents heterogeneity, and there is on average no macroscopic overlap between agents configurations ($q=0$).
In the last region $k>k_{\rm c}(T)$, the coupling between agents dominates over agents heterogeneity, and all agents are essentially in the same configuration, leading to a strong overlap ($q=1$) and to the emergence of a common characteristic, a phenomenon that has been called standardization in \cite{Jensen2019}.

\section{Conclusion}

We have extended here the model proposed in \cite{Jensen2019}, where agents have many internal configurations, but can select a specific configuration thanks to interactions, leading to simplified, or standardized agents. While Ref.~\cite{Jensen2019} focused on homogeneous agents, we have in the present work extended the model to account for agents heterogeneity, introduced through random idiosyncratic utilities associated with each configurations of each agent. Heterogeneity is introduced in a minimal way, directly inspired from the Random Energy Model for glasses \cite{DerridaPRL}.
Including heterogeneity in the model leads to the onset of a new phase at low temperature and small coupling, when heterogeneity dominates over interactions among agents.
In physical terms, this new phase shares similarity with a glass phase. The main difference here is that we consider not a single glassy system as in physics, but a large assembly of interacting agents, each of which experiencing an internal glass transition. In this glassy phase, agents are stuck in a small number of possible internal configurations, those with maximal idiosyncratic utility. In spite of the individual simplification of agents, the assembly remains heterogeneous, because all agents are in different configurations and do not share common characteristics.
Increasing the coupling strength, interactions eventually dominate over heterogeneity, effectively leading to simple agents all occupying the same configuration and sharing the same characteristic. The latter phase is similar to the low temperature phase found in the homogeneous model of \cite{Jensen2019}, where agents have been called standardized.
Finally, at high temperature, agents keep their internal complexity and do not significantly feel interactions or heterogeneity, again similarly to the homogeneous model \cite{Jensen2019}.

A further step in the study of this model would to study possible collective effects that could arise once agents are simplified. This has been done in the homogeneous model by assuming that standardized agents could be in one of two distinct states, corresponding for instance to two different opinions (or to a spin in physical terms). Interactions between these binary degrees of freedom may then lead, at low temperature, to a collectively ordered state. It would be interesting to see in more details how agents heterogeneity could possibly modify this simple picture. In addition, the role of a lower connectivity (in the present work, agents interact with all other agents) would clearly deserve to be investigated, as connectivity is known in many situations to modify critical properties \cite{LeBellac}.

\medskip

\section*{Acknowledgment}
The author is grateful to Pablo Jensen for a critical reading of the manuscript and interesting comments.

\bibliographystyle{elsarticle-num-names} 
\bibliography{biblio-srem}

\begin{thebibliography}{23}
\expandafter\ifx\csname natexlab\endcsname\relax\def\natexlab#1{#1}\fi
\providecommand{\url}[1]{\texttt{#1}}
\providecommand{\href}[2]{#2}
\providecommand{\path}[1]{#1}
\providecommand{\DOIprefix}{doi:}
\providecommand{\ArXivprefix}{arXiv:}
\providecommand{\URLprefix}{URL: }
\providecommand{\Pubmedprefix}{pmid:}
\providecommand{\doi}[1]{\href{http://dx.doi.org/#1}{\path{#1}}}
\providecommand{\Pubmed}[1]{\href{pmid:#1}{\path{#1}}}
\providecommand{\bibinfo}[2]{#2}
\ifx\xfnm\relax \def\xfnm[#1]{\unskip,\space#1}\fi
\bibitem[{Castellano et~al.(2009)Castellano, Fortunato, and
  Loreto}]{Castellano}
\bibinfo{author}{C.~Castellano}, \bibinfo{author}{S.~Fortunato},
  \bibinfo{author}{V.~Loreto},
\newblock \bibinfo{title}{Statistical physics of social dynamics},
\newblock \bibinfo{journal}{Rev Mod Phys} \bibinfo{volume}{81}
  (\bibinfo{year}{2009}) \bibinfo{pages}{591}.
\bibitem[{Bouchaud et~al.(2007)Bouchaud, M\'ezard, and Dalibard}]{LesHouches}
\bibinfo{editor}{J.-P. Bouchaud}, \bibinfo{editor}{M.~M\'ezard},
  \bibinfo{editor}{J.~Dalibard} (Eds.), \bibinfo{title}{Complex systems},
  \bibinfo{publisher}{Elsevier}, \bibinfo{year}{2007}.
\bibitem[{Barrat et~al.(2008)Barrat, Barth\'elemy, and Vespignani}]{BarratBook}
\bibinfo{author}{A.~Barrat}, \bibinfo{author}{M.~Barth\'elemy},
  \bibinfo{author}{A.~Vespignani}, \bibinfo{title}{Dynamical Processes on
  Complex Networks}, \bibinfo{publisher}{Cambridge University Press},
  \bibinfo{year}{2008}.
\bibitem[{Bouchaud(2013)}]{BouchaudJSP13}
\bibinfo{author}{J.-P. Bouchaud},
\newblock \bibinfo{title}{Crises and collective socio-economic phenomena:
  simple models and challenges},
\newblock \bibinfo{journal}{J. Stat. Phys.} \bibinfo{volume}{151}
  (\bibinfo{year}{2013}) \bibinfo{pages}{567}.
\bibitem[{Anderson(1972)}]{Anderson}
\bibinfo{author}{P.~W. Anderson},
\newblock \bibinfo{title}{More is different},
\newblock \bibinfo{journal}{Science} \bibinfo{volume}{177}
  (\bibinfo{year}{1972}) \bibinfo{pages}{393}.
\bibitem[{Chaikin and Lubensky(1995)}]{Chaikin}
\bibinfo{author}{P.~M. Chaikin}, \bibinfo{author}{T.~C. Lubensky},
  \bibinfo{title}{Principles of condensed matter physics},
  \bibinfo{publisher}{Cambridge University Press}, \bibinfo{year}{1995}.
\bibitem[{Le~Bellac(1992)}]{LeBellac}
\bibinfo{author}{M.~Le~Bellac}, \bibinfo{title}{Quantum and statistical field
  theory}, \bibinfo{publisher}{Oxford University Press}, \bibinfo{year}{1992}.
\bibitem[{Marchetti et~al.(2013)Marchetti, Joanny, Ramaswamy, Liverpool, Prost,
  Rao, and Simha}]{activematterRMP}
\bibinfo{author}{M.~C. Marchetti}, \bibinfo{author}{J.-F. Joanny},
  \bibinfo{author}{S.~Ramaswamy}, \bibinfo{author}{T.~B. Liverpool},
  \bibinfo{author}{J.~Prost}, \bibinfo{author}{M.~Rao}, \bibinfo{author}{R.~A.
  Simha},
\newblock \bibinfo{title}{Hydrodynamics of soft active matter},
\newblock \bibinfo{journal}{Reviews of Modern Physics} \bibinfo{volume}{85}
  (\bibinfo{year}{2013}) \bibinfo{pages}{1143}.
\bibitem[{de~Gennes(1999)}]{deGennesRMP}
\bibinfo{author}{P.~G. de~Gennes},
\newblock \bibinfo{title}{Granular matter: a tentative view},
\newblock \bibinfo{journal}{Rev. Mod. Phys.} \bibinfo{volume}{71}
  (\bibinfo{year}{1999}) \bibinfo{pages}{S374}.
\bibitem[{Puglisi(2015)}]{PuglisiBook}
\bibinfo{author}{A.~Puglisi}, \bibinfo{title}{Transport and fluctuations in
  granular fluids}, \bibinfo{publisher}{Springer}, \bibinfo{year}{2015}.
\bibitem[{Drossel(2001)}]{DrosselAdvPhys}
\bibinfo{author}{B.~Drossel},
\newblock \bibinfo{title}{Biological evolution and statistical physics},
\newblock \bibinfo{journal}{Adv. Phys.} \bibinfo{volume}{50}
  (\bibinfo{year}{2001}) \bibinfo{pages}{209}.
\bibitem[{Sella and Hirsh(2005)}]{SellaPNAS}
\bibinfo{author}{G.~Sella}, \bibinfo{author}{A.~E. Hirsh},
\newblock \bibinfo{title}{The application of statistical physics to
  evolutionary biology},
\newblock \bibinfo{journal}{Proc. Nat. Acad. Sci. USA} \bibinfo{volume}{102}
  (\bibinfo{year}{2005}) \bibinfo{pages}{9541}.
\bibitem[{Sznajd-Weron and Snajd(2000)}]{Sznajd}
\bibinfo{author}{K.~Sznajd-Weron}, \bibinfo{author}{J.~Snajd},
\newblock \bibinfo{title}{Opinion evolution in closed community},
\newblock \bibinfo{journal}{Int. J. Mod. Phys. C} \bibinfo{volume}{11}
  (\bibinfo{year}{2000}) \bibinfo{pages}{1157}.
\bibitem[{Deffuant et~al.(2001)Deffuant, Neau, Amblard, and
  Weisbuch}]{Deffuant}
\bibinfo{author}{G.~Deffuant}, \bibinfo{author}{D.~Neau},
  \bibinfo{author}{F.~Amblard}, \bibinfo{author}{G.~Weisbuch},
\newblock \bibinfo{title}{Mixing beliefs among interacting agents},
\newblock \bibinfo{journal}{Advances in Complex Systems} \bibinfo{volume}{3}
  (\bibinfo{year}{2001}) \bibinfo{pages}{87}.
\bibitem[{Bertin and Jensen(2019)}]{Jensen2019}
\bibinfo{author}{E.~Bertin}, \bibinfo{author}{P.~Jensen},
\newblock \bibinfo{title}{In social complex systems, the whole can be more or
  less than (the sum of) the parts},
\newblock \bibinfo{journal}{C. R. Physique} \bibinfo{volume}{20}
  (\bibinfo{year}{2019}) \bibinfo{pages}{329}.
\bibitem[{Latour et~al.(2012)Latour, Jensen, Venturini, Grauwin, and
  Boullier}]{Latour}
\bibinfo{author}{B.~Latour}, \bibinfo{author}{P.~Jensen},
  \bibinfo{author}{T.~Venturini}, \bibinfo{author}{S.~Grauwin},
  \bibinfo{author}{D.~Boullier},
\newblock \bibinfo{title}{The whole is always smaller than its parts, a digital
  test of gabriel tardes monads},
\newblock \bibinfo{journal}{The British Journal of Sociology}
  \bibinfo{volume}{63} (\bibinfo{year}{2012}) \bibinfo{pages}{590}.
\bibitem[{Jensen(2019)}]{Jensen-politics}
\bibinfo{author}{P.~Jensen},
\newblock \bibinfo{title}{The politics of physicists' social models},
\newblock \bibinfo{journal}{C. R. Physique} \bibinfo{volume}{20}
  (\bibinfo{year}{2019}) \bibinfo{pages}{380}.
\bibitem[{Phan et~al.(2004)Phan, Gordon, and Nadal}]{Nadal04}
\bibinfo{author}{D.~Phan}, \bibinfo{author}{M.~B. Gordon},
  \bibinfo{author}{J.-P. Nadal},
\newblock \bibinfo{title}{Social interactions in economic theory: an insight
  from statistical mechanics},
\newblock in: \bibinfo{editor}{J.-P. Nadal}, \bibinfo{editor}{P.~Bourgine}
  (Eds.), \bibinfo{booktitle}{Cognitive Economics},
  \bibinfo{publisher}{Springer}, \bibinfo{year}{2004}, p. \bibinfo{pages}{335}.
\bibitem[{Derrida(1980)}]{DerridaPRL}
\bibinfo{author}{B.~Derrida},
\newblock \bibinfo{title}{Random-energy model: Limit of a family of disordered
  models},
\newblock \bibinfo{journal}{Phys. Rev. Lett.} \bibinfo{volume}{45}
  (\bibinfo{year}{1980}) \bibinfo{pages}{79}.
\bibitem[{Derrida(1981)}]{DerridaPRB}
\bibinfo{author}{B.~Derrida},
\newblock \bibinfo{title}{Random-energy model: An exactly solvable model of
  disordered systems},
\newblock \bibinfo{journal}{Phys. Rev. B} \bibinfo{volume}{24}
  (\bibinfo{year}{1981}) \bibinfo{pages}{2613}.
\bibitem[{Bouchaud and M\'ezard(1997)}]{BouchMez97}
\bibinfo{author}{J.-P. Bouchaud}, \bibinfo{author}{M.~M\'ezard},
\newblock \bibinfo{title}{Universality classes for extreme-value statistics},
\newblock \bibinfo{journal}{J. Phys. A: Math. Gen.} \bibinfo{volume}{30}
  (\bibinfo{year}{1997}) \bibinfo{pages}{7997}.
\bibitem[{Wu(1982)}]{review-Potts}
\bibinfo{author}{F.~Y. Wu},
\newblock \bibinfo{title}{The potts model},
\newblock \bibinfo{journal}{Rev. Mod. Phys.} \bibinfo{volume}{54}
  (\bibinfo{year}{1982}) \bibinfo{pages}{235}.
\bibitem[{Blume et~al.(1971)Blume, Emery, and Griffiths}]{BEG71}
\bibinfo{author}{M.~Blume}, \bibinfo{author}{V.~J. Emery},
  \bibinfo{author}{R.~B. Griffiths},
\newblock \bibinfo{title}{Ising model for the $\lambda$ transition and phase
  separation in he$^3$-he$^4$},
\newblock \bibinfo{journal}{Phys. Rev. A} \bibinfo{volume}{4}
  (\bibinfo{year}{1971}) \bibinfo{pages}{1071}.

\end{thebibliography}

\end{document}